\documentclass{emulateapj}
\usepackage{graphicx}
\usepackage{bm}
\usepackage{amsmath}
\usepackage{natbib}

\newcommand{\eqb}{\begin{eqnarray}}
\newcommand{\eqe}{\end{eqnarray}}
\newcommand{\diff}{\textrm{d}}

\newcommand{\msolar}{\mbox{M$_\odot$}}
\newcommand{\ppar}{{p_\|}}

\newcommand{\pperp}{{p_\bot}}

\newcommand{\betaw}{\beta_{\rm w}}
\newcommand{\gammaw}{\gamma_{\rm w}}

\newcommand{\pparexpand[1]}{{p_\|^{(#1)}}}

\newcommand{\gammaexpand[1]}{{\gamma^{(#1)}}}
\newcommand{\nexpand[1]}{{n^{(#1)}}}
\newcommand{\Deltaexpand[1]}{{\Delta^{(#1)}}}
\newcommand{\pperpexpand[1]}{{p_\bot^{(#1)}}}
\newcommand{\Eexpand[1]}{{E^{(#1)}}}
\newcommand{\Bexpand[1]}{{B^{(#1)}}}

\newcommand{\pthetaexpand[1]}{{p_{\hat{\theta}}^{(#1)}}}
\newcommand{\pphiexpand[1]}{{p_{\hat{\varphi}}^{(#1)}}}

\slugcomment{Corrected version of ApJ 729, 104 (2011)}
\shorttitle{Charge-starved blazar jets}
\begin{document}
\title{Charge-starved, relativistic jets and blazar variability}
\author{John G. Kirk and Iwona Mochol}
\affil{Max-Planck-Institut f\"ur Kernphysik, Postfach 10~39~80,
69029 Heidelberg, Germany}
\email{john.kirk@mpi-hd.mpg.de, iwona.mochol@mpi-hd.mpg.de}

\begin{abstract}
  High energy emission from blazars is thought to arise in a
  relativistic jet launched by a supermassive black hole. The emission
  site must be far from the hole and the jet relativistic, in order to
  avoid absorption of the photons. In extreme cases, rapid variability of the
  emission suggests 
  that structures of length-scale smaller than the
  gravitational radius of the central black hole are imprinted on the 
jet as it is launched, and modulate the
  radiation released after it has been accelerated to high
  Lorentz factor. We propose a mechanism which can account for the
  acceleration of the jet, and for the rapid variability of the
  radiation, based on the propagation characteristics of large-amplitude
  waves in charge-starved, polar jets. Using a two-fluid ($e^\pm$) 
description, we find the outflows exhibit a delayed 
 acceleration phase, that starts when the inertia associated 
 with the wave currents becomes important. 
   The fluids propagate with the wave 
at approximately the sonic speed, corresponding to 
a bulk Lorentz factor $\gamma\approx10^4
  \Delta t_{100}^{1/3}\kappa_{r_{\rm g}}^{-1/3}L_{46}^{1/6}M_9^{-1/3}$
  out to radius $r_1\approx\Delta t_{100}^{1/3}
  \kappa_{r_{\rm g}}^{2/3}L_{46}^{1/6}M_9^{2/3}\, \textrm{pc}$, 
after which the Lorentz factor accelerates as  $\gamma\propto r$.  
($\Delta
  t_{100}$ is the variability time in units of $100\,$s,
  $\kappa_{r_{\rm g}}$ the pair multiplicity at one gravitational
  radius, $L_{46}$ the \lq\lq $4\pi$-luminosity\rq\rq\ of the jet in
  units of $10^{46}\,\textrm{erg/s}$, and $M_9$ the black-hole mass in
    units of $10^9\,\msolar$.) The time-structure imprinted on the jet
    at launch modulates photons produced by the
    accelerating jet provided $\kappa_{r_{\rm g}}<14\, \Delta t_{100}
    L_{46}^{1/8}M_{9}^{-1}$, suggesting that very rapid variability 
is confined to sources in which the electromagnetic cascade
in the black-hole magnetosphere is not prolific.
\end{abstract}


\section{Introduction}
Observations by the H.E.S.S. collaboration of TeV 
gamma-ray emission from the blazar PKS~2155-304
reveal very rapid variability
\citep{HESS_2155a_07,HESS_2155b_10}, at very high flux 
levels. In the most extreme flare, variations
on a timescale of a few hundred seconds at a flux level
corresponding to an isotropic luminosity of $10^{46}\,\textrm{erg/s}$
were measured. If, as expected, the mass $M$ 
of the central black hole is
$2\times10^9\,\msolar$, these observations imply
structure in the jet that is 
roughly one hundred times smaller than the gravitational radius
$r_{\rm g}=GM/c^2$ \citep{begelmanfabianrees08}. 
Though this is the most extreme example, 
several other blazars exhibit very rapid variability in 
GeV and TeV 
gamma-rays \citep[e.g.,][]{albertetal07,ackermannetal10}, which it
is proving difficult to accommodate in the standard 
synchrotron-self-Compton picture, mainly 
because of the very high Lorentz factor 
and low magnetisation required of the jet 
\citep{levinson07,boutelierhenripetrucci08,graffetal08,katarzynskietal08,kusunosetakahara08,mastichiadismoraitis08,neronovsimikozsibiryakov08,ghisellinietal09,gianniosuzdenskybegelman09,paggietal09,tammiduffy09,riegervolpe10,nalewajkoetal10}.
 
In the framework of ideal MHD, radial (uncollimated) relativistic jets 
do not accelerate after they pass through
the fast magnetosonic point.
Collimation, however, requires special boundary conditions 
\citep{lyubarsky09,lyubarsky10}. It is, therefore, difficult
to envisage the production of a jet with high Lorentz factor 
and low magnetisation. An isolated, impulsive, 
ejection event in an ideal MHD 
flow is able to circumvent this problem \citep{granotkomissarovspitkovsky10}, 
but, in a fluctuating jet, 
dispersion filters out the small timescale structure,
and limits the acceleration \citep{levinson10}.
It appears, therefore, that it may be necessary to go beyond the ideal
MHD approximation in order to understand the observations of
very rapid variability in blazars. 

Non-ideal MHD effects can become important when plasma is
{\em charge-starved}:
If the number density of charged particles is limited, 
this places a maximum on the absolute value of the charge density, which 
can, for example, result
in the inability of the plasma 
to screen out the component of the electric field parallel to the 
magnetic field. 
Charge starvation also places an upper limit on the available
current density. As this limit is approached, the relative drift-speed 
of the charged components becomes relativistic, and their associated inertia
begins to contribute to the fluid stress-energy tensor. 
This situation might plausibly arise in a black-hole 
magnetosphere, because axisymmetric, general-relativistic, 
MHD simulations of jets launched
by accreting, 
rotating black holes \citep{devilliershawley03,mckinneygammie04}
reveal a conical region around the rotation axis into which the accreting
plasma does not penetrate. As in a pulsar magnetosphere, the matter
density in this region is likely to be determined not by accretion, but
by the rate
at which electron-positron pairs are created in the strong
electromagnetic fields that penetrate 
it \citep{goldreichjulian69,blandfordznajek77,levinson00}. 
If an outflow results, the 
density of pairs decreases, and,
far from the hole,
non-MHD effects connected with 
particle inertia can become important. 

In the case of an axisymmetric, force-free magnetosphere,
it is known that plasma is ejected, carrying off energy
mainly in the form of Poynting flux via the \lq\lq
Blandford-Znajek\rq\rq\ mechanism. On the axis itself, the energy flux
vanishes, so that the polar regions of the jet do not dominate the 
overall energetics. However, observations of rapid variability suggest
that axisymmetry may not be a good approximation, since they imply 
small-scale structure in the black-hole magnetosphere. 
Simulations in which the black-hole spin and the asymptotic magnetic field
are misaligned find a Poynting flux comparable to the 
Blandford-Znajek value \citep{palenzuelaetal10}, 
making it plausible that also more complex
non-axisymmetric field structures can power a substantial,
magnetically dominated, polar jet. 

In the following, we develop this idea by examining the 
propagation characteristics
of nonlinear electromagnetic 
waves above the polar regions of a rotating black hole. Using 
a model consisting of cold electron and positron fluids, we show that a 
circularly polarised magnetic
shear propagates radially outwards at roughly the fast magnetosonic speed
until it reaches the point where the ideal MHD description loses its validity.
In (non-ideal)~MHD language, 
this happens when the inertia of the plasma particles begins to affect
the conductivity, i.e., when the drift speeds implied by the 
plasma current become relativistic
\citep{melatosmelrose96,meier04}. This can occur at a large distance
from the black hole, depending on the density of injected pairs.  
The wave then goes through a phase of delayed acceleration
in which it converts Poynting flux into kinetic energy flux.  
If particles radiate gamma-rays in the acceleration zone,
then, despite the large spatial extent of the source, 
the spatio-temporal structure of the 
shear wave, that is imprinted on it close to the black hole, modulates 
the radiation, provided the mass-loading of the jet is 
sufficiently small. 

In section~\ref{parameters} we discuss the parameters used to specify
the physical conditions in the jet. The two-fluid jet model is presented in 
section~\ref{two-fluid}. First, the nonlinear 
plane-wave solution representing a magnetic
shear is derived, then radial
propagation in spherical geometry is discussed. It is shown in the 
Appendix that general relativistic effects drop out in 
the short-wavelength approximation ($c/\omega\ll r$) when the Kerr metric 
is used. The application to blazar variability is discussed in 
section~\ref{blazars}, and our conclusions summarised in \ref{conclusions}.

\section{Jet parameters}
\label{parameters}
Consider a radial outflow consisting of an electron-positron 
plasma emerging from the polar regions of the magnetosphere of 
a rotating black hole, and denote by $L$ and $\dot{M}$
the total luminosity and mass-flux
carried in a solid opening angle $\Omega_{\rm s}$.
The physical conditions in the flow can be specified via three
dimensionless parameters:
\begin{enumerate}
\item
The nonlinearity or strength parameter $a$ is a dimensionless 
measure of the energy-flux density.
For a circularly polarized 
vacuum electromagnetic wave of frequency $\omega/2\pi$
and electric field $E$, the strength parameter
is conventionally defined as $a=eE/(mc\omega)$, and 
measures the Lorentz factor that an electron would achieve
if it were accelerated from rest over a distance  of 
$\left(2\pi\right)^{-1}$ times one wavelength in the field $E$.
Assuming radial propagation, the corresponding luminosity is
\eqb
L&=&\frac{m^2 c^3\omega^2 a^2\Omega_{\rm s}r^2}{4\pi e^2}
\enspace,
\eqe
and we use this expression to define $a$ for a 
general (non-vacuum) wave.
In the absence of 
radiation losses, $a\propto r^{-1}$, and
is determined by specifying its value at some fiducial radius. 
For this we chose $r_0=c/\omega$, 
although our treatment is, of course, valid only for $r\gg r_0$ and we 
certainly do not expect radial flow to extend to such small radii. 
With this choice, $a_0$ is independent of $\omega$:
\eqb
a&=&a_0\left(r_0/r\right)
\\
a_0&=&\left[\frac{4\pi e^2 L}{m^2c^5\Omega_{\rm s}}\right]^{1/2}
\nonumber\\
&=&3.4\times10^{14} L_{46}^{1/2}
\enspace,
\label{azerodef}
\eqe
where $L_{46}=\left(4\pi/\Omega_{\rm s}\right) 
L/\left(10^{46}\,\textrm{erg/s}\right)$ is the 
\lq\lq isotropic\rq\rq\ or \lq\lq$4\pi$\rq\rq\ luminosity 
of the jet, scaled appropriately.
The energy {\em radiated} per unit solid angle by a jet is directly measurable
if the distance to the object is known. 
For the rapidly variable gamma-ray flare of PKS~2155-304, it corresponds to 
an isotropic luminosity of roughly $10^{46}\,\textrm{erg/s}$, 
so that, for this object, $L_{46}\gtrsim 1$.
\item
The mass-loading of the wind is conventionally described by the $\mu$-parameter
introduced by \citet{michel69}:
\eqb
\mu&=&L/\dot{M}c^2
\enspace.
\eqe
In the case of an electron-positron jet, $\mu$ denotes the Lorentz factor
each particle would have if the entire luminosity was carried by 
a cold, unmagnetised flow. It is constant in those parts of the jet in which 
pair creation and radiation losses can be neglected. 
\item
The magnetisation parameter $\sigma$ describes the ratio of 
the energy flux carried by electromagnetic fields to that 
carried by particles. For monoenergetic electrons and positrons
of Lorentz factor $\gamma$, 
\eqb
\sigma&=&\left(\mu/\gamma\right)-1
\enspace.
\eqe
In a cold, non-accelerating, ideal MHD flow, $\sigma$ is constant
(assuming pair creation and radiation losses are negligible).
However, as we show below, $\sigma$ is not constant in 
charge-starved jets, even in the absence of dissipation. For this reason,
we specify its value at the \lq\lq launching\rq\rq\ radius, inside of which
the ideal MHD approximation is assumed to hold, and denote this quantity
by $\sigma_0$, even though the region of 
constant $\sigma$ is unlikely to extend to radii as small as 
$r\sim c/\omega$. The particle Lorentz factor at
the launching point is then
\eqb
\gamma_0&=&\mu/\left(\sigma_0+1\right)
\enspace.
\eqe
\end{enumerate}

The mass-loading parameter $\mu$ is determined by the physics of the 
pair-production cascade close to the black
hole. A more intuitive measure, therefore, is the pair multiplicity
$\kappa$, which relates the pair (proper)
number density $n_\pm$ to the number density of 
electrons (or positrons) needed to screen out the (magnetic-)field-aligned
component of the rotation-induced electric field. 
Adopting the definition conventionally used in pulsar physics, but 
replacing the angular velocity of the neutron star by $c/r_{\rm g}$ gives
\citep[e.g.,][]{lyubarskykirk01}
\eqb
\kappa&=&\gamma_\pm n_\pm\left(\frac{B}{2\pi e r_{\rm g}}\right)^{-1}
\enspace,
\eqe
where $\gamma_\pm$ is the Lorentz factor of the fluids.
In the inner regions of the flows we consider, where 
$\sigma=\sigma_0\gg1$,  
the fluids move non-relativistically in the 
wave frame, so that
$\kappa
\approx a/\left(4\mu\right)$.
Thus, in the absence of radiation losses and pair production, 
$\kappa\propto r^{-1}$ in this region. 
Physically, it is the value of $\kappa$ at the outer boundary
of the pair-production region that is most relevant. This is thought to be 
close to the black hole, but its precise location
is unknown. In the following, therefore, we specify $\kappa$ by its value 
$\kappa_{r_{\rm g}}$ at 
$r=r_{\rm g}$:
\eqb
\kappa_{r_{\rm g}}&\approx& \frac{a_0}{4\mu}
\left(\frac{c}{\omega r_{\rm g}}\right)
\enspace.
\eqe

\section{The two-fluid model}
\label{two-fluid}
The simplest model of an electron-positron plasma that 
captures the physics connected with the finite inertia of the 
charge-carriers is that of two cold, oppositely charged fluids
(denoted by suffices $-$ and $+$). We adopt this model, 
embed it in a Kerr metric, following \citet{khanna98} and 
\citet{koide09},
and look for large-amplitude waves propagating
in the radial direction in Boyer-Lindquist coordinates. 
To keep the analysis tractable, only 
transverse waves 
with vanishing phase-averaged components of the electric and 
magnetic fields are treated. 
In this case, since $\nabla\cdot\bm{E}=0$, the number 
density is the same for each fluid. 
We further assume the wave carries no radial current, so that the
radial fluid velocities also equal each other: 
$v_{\hat{r}+}=v_{\hat{r}-}=v_{\hat{r}}$, and restrict the treatment to waves in which
the meridional and azimuthal fluid velocities are equal in magnitude but of 
opposite sign: $v_{\hat{\theta}+}=-v_{\hat{\theta}-}$, $v_{\hat{\phi}+}=-v_{\hat{\phi}-}$. 
It follows that the fluids have the same Lorentz factor:
$\gamma_+=\gamma_-=\gamma=c/\left(c^2-v_{\hat{r}}^2-v_{\hat{\theta}+}^2-v_{\hat{\phi}+}^2\right)^{1/2}$,
and the same radial component of the four-velocity, which we write
as a dimensionless momentum: $\ppar=v_{\hat{r}}\gamma/c$. 
Circularly polarised waves are likely to be the most important
in the polar regions of a rotating black hole, and these are best treated
by introducing complex quantities to describe the transverse 
components of the fluid momenta:
$\pperp=\left(v_{\hat{\theta}+}+iv_{\hat{\phi}+}\right)\gamma/c$
($=-\left(v_{\hat{\theta}-}+iv_{\hat{\phi}-}\right)\gamma/c$) and the 
electric and magnetic fields:
$E=E_{\hat{\theta}}+iE_{\hat{\phi}}$, $B=B_{\hat{\theta}}+iB_{\hat{\phi}}$. 
In Appendix~\ref{app1} we derive the continuity equation, the equations
of motion of the fluids and the two relevant Maxwell equations
(Faraday's law and Amp\`ere's law) in the small-wavelength approximation,
$r\gg c/\omega$, starting from the formulation given by \citet{khanna98}.
In the lowest order, we search for large-amplitude
plane-wave solutions using the approach introduced by 
\citet{akhiezerpolovin56}.

\subsection{Nonlinear waves}
Expressing all quantities in terms of the phase 
$\phi$ defined in (\ref{phasedef}), 
the continuity equation (\ref{contzero}) and 
Faraday's law (\ref{faradayzero}) integrate
immediately to give
\eqb
n\Delta&=&\textrm{constant}
\label{contzeroint}
\\
B&=&iE/\betaw
\enspace,
\label{faradayzeroint}
\eqe
where $\Delta=\gamma-\left(\ppar/\beta_{\rm w}\right)$, and 
$c\betaw$ is the phase speed
of the wave. The equations of motion 
to this order, (\ref{pparzero}) and (\ref{pperpzero}), are
\eqb
\omega\Delta\frac{\diff \ppar}{\diff\phi}&=&-\frac{e}{mc}\textrm{Im}
\left(\pperp B^*\right)
\label{eqmotparzero}
\\
\omega\Delta\frac{\diff \pperp}{\diff\phi}&=&-i\frac{eB}{mc}\Delta
\enspace.
\label{eqmotperpzero}
\eqe
Solutions to these equations can be found 
with superluminal phase speed, $\betaw>1$, but these waves   
do not propagate close to the black hole \citep{kirk10}. 
Here, we concentrate on subluminal waves, for 
which the condition $\Delta=0$ holds. Physically, this means that the 
particles are in resonance with the wave, i.e., 
the radial components of the fluid velocities equal
the phase velocity of the wave. In this case (\ref{contzeroint}) shows that 
the phase dependence of the density is arbitrary, (\ref{eqmotperpzero}) is 
trivially satisfied, and (\ref{eqmotparzero}) requires 
that the transverse components of the fluid velocities are parallel to the 
magnetic field: $\textrm{Im}\left(\pperp B^*\right)=0$. 
Thus, the plasma current is directed along
the magnetic field and, to this order, the forces exerted on the fluids
by the fields vanish. Finally, the current and magnetic field are linked 
by Ampere's equation, which, to this order, reads
\eqb
\frac{\partial B}{\partial\phi}
&=&
\frac{8\pi i n e c\pperp\betaw}{\omega\left(1-\betaw^2\right)}
\enspace.
\label{ampere2}
\eqe
Combined with (\ref{eqmotparzero}) this implies that the magnitude of the 
magnetic field is phase-independent. 
Viewed from a frame that moves radially with speed $\betaw$,
the electric
field vanishes, and the wave is 
simply a static magnetic field of constant magnitude whose direction 
rotates through $2\pi$ radians over one wavelength.
At each point, the 
current and, hence $\pperp$, is parallel to the magnetic field to zeroth order.
The rate at which the $B$-vector rotates is arbitrary, being determined by the 
dependence of the fluid density $n$ on phase. In the following, 
we select the simplest case, where
$n$, $\left|B\right|^2$ and $\left|\pperp\right|^2$ are all constant
and the wave is a monochromatic magnetic shear: 
$B\propto \pperp\propto \textrm{e}^{\pm i\phi}$.

\subsection{Radial evolution of a magnetic shear}
\label{shearevolution}
The slow evolution of the subluminal magnetic shear wave 
as it propagates outwards
at $r\gg r_{\rm g}$ is governed by the first-order equations 
in the expansion in $\epsilon\sim c/\left(\omega r\right)$, 
as derived in Appendix~\ref{app1}.
Making obvious simplifications to 
the notation, these are the continuity equation
(\ref{contfinal2}):
\eqb
\ppar&=&\mu\hat{\omega}^2/R^2
\enspace,
\label{finalset1}
\eqe
the equation of energy flux 
conservation:
\eqb
\mu&=&\gamma(1+\sigma)
\eqe
and the radial momentum equation
\eqb
\frac{d\nu}{dR}=\frac{R|\pperp|^2}{\mu\hat{\omega}^2}\enspace,
\eqe
where the momentum-flux density per unit mass-flux
\eqb
\nu&=&\ppar\left(1+\frac{1+\betaw^2}{2\betaw^2}\sigma\right)
%\enspace
\eqe
and $\hat{\omega}=\omega/\omega_{\rm p}$ (with 
the plasma frequency defined using the proper fluid density:
$\omega_{\rm p}^2=8\pi n e^2/m$), 
$R=\left(\omega r/c\right)\left(\mu/a_0\right)$ is the radius in units of 
the critical radius, inside of which the superluminal 
modes do not propagate \citep{kirk10}. 
Note that in a non-monochromatic
wave, the quantities $\ppar$, $\hat{\omega}$, $\sigma$ 
and $\gamma$ are replaced 
in these equations by their phase-averages.

According to (\ref{ampere2}), $\sigma$, as defined in (\ref{sigmadef}) 
is related to 
the fluid momentum components through
\eqb
\sigma&=&\frac{\betaw^4\gammaw^4\left|\pperp\right|^2}{\hat{\omega}^2\ppar^2}
\enspace,
\eqe
where $\gammaw=\left(1-\betaw^2\right)^{-1/2}$. The 
condition that the wave velocity equals the radial component 
of the fluid velocity, $\Delta=0$, implies
\eqb
\gammaw^2&=&\gamma^2/\left(1+\left|\pperp\right|^2\right)
\enspace.
\label{finalset2}
\eqe

The five equations (\ref{finalset1})--(\ref{finalset2}), together with
the definition
$\gamma=\left(1+\ppar^2+\left|\pperp\right|^2\right)^{1/2}$, determine
the radial dependence of the six unknown wave variables $\gammaw$,
$\gamma$, $\ppar$, $\left|\pperp\right|$, $\hat{\omega}$, and
$\sigma$. It is straightforward to reduce these to a first-order ordinary
differential equation for $\pperp(\gamma_{\rm w})$, for example. 
Solutions extend from $R=0$ to $R=\infty$ provided they are 
launched at super-magnetosonic speed: $\sigma=(\mu/\gamma)-1<\mu^{2/3}$.
At $R\rightarrow0$, 
$\left|\pperp\right|\rightarrow0$ and $\gamma\rightarrow\gammaw$.
whereas at $R\rightarrow\infty$, $\sigma\rightarrow0$ and 
$\gamma\rightarrow\mu$, so that the
wave converts all of the Poynting flux to kinetic energy at large
radius. 

\begin{figure}
\epsscale{1.}
\plotone{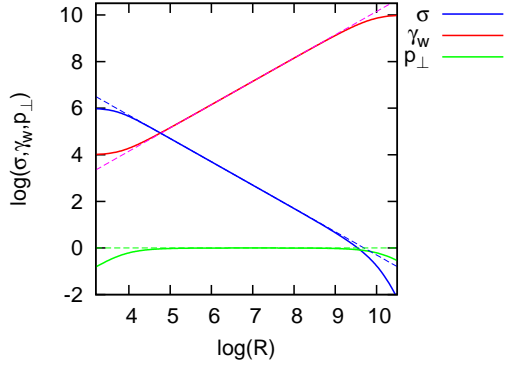}
\caption{\label{fig1}%
The magnetisation parameter $\sigma$, transverse fluid 
momentum $\pperp$ and Lorentz factor of the wave, 
$\gamma_{\rm w}$, as
functions of the dimensionless radius $R=\mu r\omega/\left(a_0 c\right)$,
for $\sigma_0=10^6$ and $\mu=10^{10}$. The approximate solutions given in Eq.~(\ref{approximatesols}) are also shown.}
\end{figure}

The radial dependence of the wave goes through three phases. At small
$R$, $\left|\pperp\right|\ll1$ and the wave is essentially a cold MHD
  structure in which the inertia associated with the current is
  negligible. There is no acceleration of either the wave speed or the
  fluids in this regime, and the magnetisation parameter $\sigma$
  remains constant at its initial value $\sigma_0$. 
Assuming $\sigma_0\gg1$, this region
is restricted to $R\ll \mu/\sigma_0$.
At intermediate radii, one readily finds an approximate solution:
\eqb
\begin{array}{r@{\,\approx\,}l@{\qquad}r@{\,\approx\,}l}
\pperp&1&
\sigma&\mu/\left(2R\right)\\
\gamma&2R&
\gamma_{\rm w}&\gamma/\sqrt{2}
\label{approximatesols}
\enspace,
\end{array}
\eqe
valid in the range 
\eqb
\mu/\sigma_0\ll R\ll \mu
\enspace.
\eqe
Finally, at large radius, $R\gg\mu$, 
only kinetic energy remains: 
$\gamma\approx\gammaw\approx\mu$, 
$\left|\pperp\right|\ll1$.
This behaviour is illustrated in Fig.~\ref{fig1}.

\section{Application to blazar variability}
\label{blazars}
The locations of the three phases of wave propagation illustrated in 
Fig.~\ref{fig1} depend on 
the parameters $a_0$, $\mu$ (or $\kappa_{r_{\rm g}}$),
$\sigma_0$ and $\omega$. As discussed in Sect.~\ref{parameters}, 
it is possible to infer values for $a_0$ and $\omega$ directly from 
the observed flux and variability timescale. 
Another parameter may be eliminated by 
fixing the wave speed at its launching point. A very slow, 
sub-magnetosonic outflow
can be described by the force-free MHD equations, and would accelerate
as  $\gamma\propto r$ \citep{buckley77}, until it approached the sonic speed,
where $\gamma_0\approx\sigma_0^{1/2}\approx\mu^{1/3}$.
On the other hand, all waves launched at super-magnetosonic speeds
($\sigma_0< \mu^{2/3}$) behave similarly, as described in 
Sect.~\ref{shearevolution},
with the acceleration
phase moving out to larger radius as the initial magnetisation
decreases. It suffices, therefore, to analyse the case of mildly
supersonic launch: $\sigma_0\approx\mu^{2/3}$, corresponding to the 
maximum magnetisation of a supersonic flow.

However, the uncertainty associated
with the unknown mass-loading of the flow can be removed only by
modelling the pair cascade. Leaving this quantity as a free parameter,
the radius $r_{\rm acc}$ at which acceleration begins, corresponding to 
$R\approx \mu/\sigma_0$, is
\eqb
r_{\rm acc}&\approx&r_{\rm g}a_0^{1/3}\kappa_{r_{\rm g}}^{2/3}
\left(\omega r_{\rm g}/c\right)^{-1/3}
\nonumber\\
&=&1.2\,\Delta t_{100}^{1/3}\kappa_{r_{\rm g}}^{2/3}
L_{46}^{1/6}M_9^{2/3}\,\textrm{pc}
\enspace,
\label{defracc}
\eqe
where we define the variation timescale in units of $100\,$s to be
$\Delta_{100}=\left(2\pi/\omega\right)/\left(100\,\textrm{s}\right)$,
and write the mass of the black hole as $M=M_9\times10^9\,\msolar$. For 
$r<r_{\rm acc}$ the Lorentz factor of the flow remains constant at roughly
the sonic speed: 
\eqb
\gamma_0&\approx&6.5\times10^3
\Delta t_{100}^{1/3}\kappa_{r_{\rm g}}^{-1/3}
L_{46}^{1/6}M_9^{-1/3}
\label{gamma0eq}
\eqe
and, for $r>r_{\rm acc}$, the Lorentz factors increase linearly
with $r$:
\eqb
\gamma_{\rm w}&\approx&\gamma/\sqrt{2}\,\approx\,7.4\times 10^3 
\left(r/1\,\textrm{pc}\right)\kappa_{r_{\rm g}}^{-1}M_9^{-1}
\enspace.
\eqe

The solutions presented in section~\ref{shearevolution}
eventually convert all of the Poynting flux to kinetic energy flux
at large radius. However,
in the case of blazars, this is unlikely to be realised, since the 
resulting Lorentz factor ($=\mu$) is very large. 
Instead, dissipative processes so far neglected, 
such as instabilities in the wave-solution, 
or interaction of the jet with the external medium, or with ambient
photons, are likely to intervene.

The wave propagates radially with fixed frequency. 
However, if it converts part of its energy into 
high-frequency ($\gg\omega$), forwardly beamed 
photons via an emissivity that 
is modulated by the wave-phase, then the difference between photon
and wave propagation speeds will lead to a smoothing of the modulation
in the observed photon signal. This loss of short-timescale variability
becomes more effective as 
the size of the radiating section of the wave increases.  
Similarly, if the photons do not propagate
exactly in the radial direction, smoothing of the modulation will be produced
by the difference in light-travel time to the observer from different 
parts of the spherical wavefront. It is straightforward to derive 
a criterion on the size of the emitting region (assumed $\sim r$) 
and the Lorentz factor of the jet, such that fluctuations of 
frequency $\omega$ are not suppressed in the 
photon signal \citep{michel71,arons79,kirketal02}:
\eqb
\gamma_{\rm w}^2 2\pi c/\omega &>& r
\enspace.
\eqe 

In the acceleration region, $\gamma_{\rm w}\propto r$, so that 
this condition is fulfilled everywhere within this region, provided 
it is satisfied at the beginning, where $r=r_{\rm acc}$, and $\gamma_{\rm w}=\gamma_0$.
Combining (\ref{defracc}) and (\ref{gamma0eq}) the requirement that
modulation on a timescale of $100\Delta t_{100}\,$seconds should not be
filtered out leads to 
an upper limit on the multiplicity:
\eqb
\kappa_{r_{\rm g}}&<& 14\, \Delta t_{100}
    L_{46}^{1/8}M_{9}^{-1}
\label{limit}
\enspace.
\eqe

Equation~(\ref{limit}) implies that electron-positron pair creation is much
less effective in the central engine of a rapidly variable blazar
than it is in a pulsar magnetosphere 
\citep[e.g.,][]{medinlai10}, but this is perhaps not 
unexpected, given that a neutron star surface is able to anchor a 
very strong magnetic
field.  
However, it also implies 
that blazars exhibiting extreme variability contain a
charge-starved magnetosphere able to support a vacuum gap
\citep{levinson00,levinson10b}. This scenario is particularly 
attractive, because the non-stationary nature of gap discharges found
in pulsar-related studies \citep{levinsonetal05,timokhin10}, suggests 
a natural source of short-timescale ($<r_{\rm g}/c$) variability in the 
outflow from a black-hole magnetosphere.
 
\section{Summary and conclusions}
\label{conclusions}
In this paper, we describe a mechanism that causes a 
magnetically dominated, radial outflow from a black-hole magnetosphere
to enter a delayed acceleration phase, starting at a 
distance from the hole given by (\ref{defracc}). Applying this
mechanism to blazar jets, we derive a constraint, (\ref{limit}),
on the pair density in the magnetosphere 
that would allow radiation produced where the jet accelerates
to retain any short-timescale structure imposed on it close to 
the launching site. 

The mechanism is based on an analysis of the propagation characteristics of 
a nonlinear wave -- specifically a circularly polarised magnetic shear -- 
in a low-density plasma. Such a wave, we suggest, is likely to be launched
in the polar regions of a rotating, accreting black hole, and, in a 
non-axisymmetric picture, may fluctuate
on a time shorter than $r_{\rm g}/c$, as indicated by observations of the source
PKS~2155-304. 
Acceleration is a result of charge-starvation -- a 
non-MHD effect that arises when the relative drift-speed of the oppositely
charged constituents in a low-density 
plasma becomes relativistic. The analysis employs
a cold two-fluid model of the plasma, and uses
a short-wavelength perturbation expansion to find the evolution of the
radially propagating, nonlinear wave. The equations are 
derived in Kerr geometry. However, under the conditions we envisage, where
the wavelength of the oscillation is of the same order in the expansion 
parameter as the gravitational radius, 
general relativistic effects do not appear
in the governing equations.

Several important problems remain to be investigated. 
These include the nature 
of the dissipation and radiation mechanisms, and the effect these might have
on the propagation of the wave, as well as the possibility of 
modelling the multi-wavelength blazar spectrum. 
Furthermore, although the picture of a circularly
polarised magnetic shear that is static in the jet frame is intuitively 
attractive, this is only one specific, nonlinear solution of the governing
equations; other polarisations and other 
modes, such as the linearly polarised \lq\lq striped wind\rq\rq\ 
\citep{lyubarskykirk01} or 
the electromagnetic mode of superluminal phase-speed \citep{kirk10}
may also prove important. Nevertheless, the underlying physical cause of the 
acceleration --- the inertia of the charge-carriers --- 
suggests that delayed jet-acceleration may be a generic phenomenon.

%\onecolumn
\appendix
\section{Equations of wave propagation in Kerr geometry}
\label{app1}
Here we derive the equations governing the radial propagation of
transverse, circularly polarised, electromagnetic waves in a plasma
consisting of cold electron and positron fluids that are embedded in a
Kerr metric. We use a short-wavelength approximation:
$c/\omega r\sim\epsilon\ll 1$, where $\omega$ is 
the wave frequency, and assume the gravitational radius is
of the same order in $\epsilon$ as the wavelength. In this appendix 
we set $G=c=1$, so that with $M$ the black-hole mass, 
$M \sim 1/\omega \sim \epsilon r$.
We start from the two-fluid equations as given by \citet{khanna98}, 
and use the (essentially standard) notation of that paper.

As measured by a fiducial observer corotating with a black hole,
the fluid 3-velocities and Lorentz factors, 
and the electric and magnetic fields are
denoted by $\bm{v}_\pm$, $\gamma_\pm$, and $\bm{E}$ and $\bm{B}$
respectively, the suffices $+$ and $-$ denoting the positron and
electron components (charge $\pm e$, mass $m$). 
The continuity equation \citep[][Eq.~(22)]{khanna98} reads
\eqb 
\left(\partial_t-\bm{\beta}\cdot\nabla\right)n_\pm \gamma_\pm
+\nabla\cdot\left(\alpha n_\pm \gamma_\pm \bm{v}_\pm\right)&=&0 
\enspace,
\label{continuity}
\eqe
where $\nabla$ is an operator in curved 3-dim space 
described by $g_{ij}$, $\alpha$ is the lapse function and 
$\bm{\beta}$ the gravitomagnetic potential.
The equations
of momentum conservation for each fluid 
\cite[][Eq.~(23)]{khanna98}
\eqb 
\alpha^{-1}\left(\partial_t-\bm{\beta}\cdot\nabla\right)\bm{S}_\pm
&=&
\varepsilon \bm{g}+
\overleftrightarrow{H}\cdot\bm{S}_\pm
-\alpha^{-1}\nabla\cdot\left(\alpha\overleftrightarrow{T}_\pm\right)
\nonumber\\
&&\pm e \gamma_\pm n_\pm \bm{E}
\pm e \gamma_\pm n_\pm\bm{v}_\pm \times\bm{B} 
\eqe
reduce to the equations of motion 
\eqb 
\frac{1}{\alpha}\frac{\partial}{\partial t}
+\left(\bm{v}_\pm-\frac{\bm{\beta}}{\alpha}\right)\cdot\bm{\nabla}
p^i_\pm
&=&
\gamma_\pm g^i+H^{ij}p_{j\pm}
\nonumber\\
&&\pm\frac{e}{m}\left(E^i+\epsilon^{ijk}v_{j\pm}B_k\right) 
\label{motion}
\eqe
in the special case of cold, collisionless fluids
(in Khanna's notation $\varepsilon=\gamma^2_\pm n_\pm m$, 
$\bm{S}_\pm=\gamma^2_\pm n_\pm m\bm{v}_\pm$ 
and $\overleftrightarrow{T}=n_\pm m\gamma^2_\pm\bm{v}_\pm\bm{v}_\pm$).
Faraday's law and Ampere's law are 
\eqb
\nabla\times(\alpha\bm{E})
&=&
-\left(\partial_t-\mathcal{L}_{\bm{\beta}}\right)\bm{B} 
\label{faraday}
\\
\nabla\times\left(\alpha\bm{B}\right)
&=&
\left(\partial_t-\mathcal{L}_{\bm{\beta}}\right)\bm{E}
+4\pi\alpha e\left(n_+\bm{p}_+ - n_-\bm{p}_-\right) 
\enspace,
\label{ampere}
\eqe
where $\mathcal{L}_{\bm{\beta}}$ is a Lie derivative along $\bm{\beta}$:
\eqb 
\mathcal{L}_{\bm{\beta}}\bm{E}
&=&
\left(\bm{\beta}\cdot\nabla\right)\bm{E}
-\left(\bm{E}\cdot\nabla\right)\bm{\beta} 
\enspace.
\eqe

We now restrict the treatment to radially propagating waves by 
assuming the field and fluid variables depend only on $t$ and $r$,
and introduce the wave phase $\phi$, a function of $t$ and $r$ that
depends on the (as yet unspecified) wave phase velocity
$\betaw(r)$, which is a function of $r$ alone. 
The phase 
of an outwardly propagating vacuum wave in Kerr geometry is 
$\phi_{\rm vac}=\omega\left(t-r_*\right)$, where
\eqb 
r_{*}
&=&
\int^r \frac{r'^2+a^2}{r'^2-2Mr'+a^2}\diff r' 
\enspace,
\eqe
$a$ (in this appendix) is the Kerr parameter, 
and $\omega$ is the wave frequency measured by an observer at infinity
\citep[e.g.,][Eq~(8.66]{thornezurekprice86}. In analogy with this expression
we write the phase of the nonlinear wave as 
\eqb 
\phi &=& 
\omega\left(t-\int^r\frac{r'^{2}+a^2}
{\left(r'^2-2Mr'+a^2\right)\betaw(r')}\diff r'\right)
\enspace. 
\label{phasedef}
\eqe
Because we restrict the treatment to transverse waves, $\bm{E}$ and 
$\bm{B}$ are automatically divergence-free, and the plasma is 
charge-neutral. The waves of interest have
$v_{\hat{r}+}=v_{\hat{r}-}=v_{\hat{r}}$ and $v_{\hat{\theta}+}=-v_{\hat{\theta}-}$, 
$v_{\hat{\phi}+}=-v_{\hat{\phi}-}$, so that 
$\gamma_+=\gamma_-=\gamma$, $n_+=n_-=n$, and we describe them using 
complex quantities for the transverse (dimensionless) momenta 
$\pperp=\gamma\left(v_{\hat{\theta}+}+iv_{\hat{\phi}+}\right)$ and 
for the electric and magnetic
fields: $E=E_{\hat{\theta}}+iE_{\hat{\phi}}$, $B=B_{\hat{\theta}}+iB_{\hat{\phi}}$. In accordance 
with standard notation, we rename the radial component 
$\ppar=p_{\hat{r}}$. 

Transforming the 
independent variables in (\ref{continuity}--\ref{ampere}) from
$(t,r)$ into a \lq\lq fast\rq\rq\ phase variable and a slow
radial coordinate: $(\phi,\rho)$, where $\rho=\epsilon r$, we now expand 
in the small parameter $\epsilon$, assuming $\rho\sim r_{\rm g}$. 
Keeping terms of zeroth and
first order, the derivatives are replaced 
according
to 
\eqb
\frac{\partial}{\partial t} 
&\rightarrow& \omega\frac{\partial}{\partial\phi} 
\\
\frac{\partial}{\partial r} 
&\rightarrow& 
\frac{\partial {\rho}}{\partial r}\frac{\partial}{\partial {\rho}}+\frac{\partial \phi}{\partial r}\frac{\partial}{\partial \phi} 
\nonumber\\
&=& 
\epsilon\frac{\partial}{\partial {\rho}} -\left(1+\epsilon\frac{2M}{{\rho}}\right) \frac{\omega}{\betaw}\frac{\partial}{\partial \phi} 
\\
\frac{\partial}{\partial t}
+\left(\alpha\bm{v}_\pm-\bm{\beta}\right)\cdot\bm{\nabla}
&\rightarrow& 
\epsilon v_{\hat{r},\pm}\frac{\partial}{\partial {\rho}}
\nonumber\\
&&+\omega\left(1-\frac{v_{\hat{r},\pm}}{\betaw}\right)\frac{\partial}{\partial \phi}
\enspace.
\eqe

Expanding the dependent variables according to 
$\ppar=\pparexpand[0]+\epsilon\pparexpand[1]$ etc., one finds the zeroth-order
equations are those of continuity:
\eqb
\omega\frac{\partial}{\partial\phi}\left(\nexpand[0]\Deltaexpand[0]\right)
&=&0 
\enspace,
\label{contzero} 
\eqe
Faraday's and Amp\`ere's laws:
\eqb
-\frac{\omega}{\betaw}\frac{\partial \Eexpand[0]}{\partial\phi}
-i\omega\frac{\partial \Bexpand[0]}{\partial\phi}
&=&0 \label{faradayzero} \\
-\frac{\omega}{\betaw}\frac{\partial \Bexpand[0]}{\partial\phi}
+i\omega\frac{\partial \Eexpand[0]}{\partial\phi}+i8\pi e \nexpand[0]\pperpexpand[0]
&=&0 
\enspace,
\label{amperezero} 
\eqe
and momentum/energy conservation:
\eqb
\omega\Deltaexpand[0]\frac{\partial \pparexpand[0]}{\partial\phi}
+\frac{e}{m}\textrm{Im}\left(\pperpexpand[0] \Bexpand[0]^*\right)
&=&0 
\label{pparzero}
\\
\omega\Deltaexpand[0]\frac{\partial \pperpexpand[0]}{\partial\phi}
-\frac{e}{m}\left(\gammaexpand[0] \Eexpand[0]+i\pparexpand[0] \Bexpand[0]\right)
&=&0 
\label{pperpzero}
\\
\omega\Deltaexpand[0]\frac{\partial \gammaexpand[0]}{\partial\phi}
-\frac{e}{m}\textrm{Re}\left(\pperpexpand[0] \Eexpand[0]^*\right)
&=&0
\enspace,
\label{gammazero}
\eqe
where $\Delta=\gamma-\ppar/\betaw$. The monochromatic, 
subluminal solution to these equations
has $\Deltaexpand[0]=0$, and $\left|\Bexpand[0]\right|^2$, 
$\left|\Eexpand[0]\right|^2$, 
$\left|\pperpexpand[0]\right|^2$, $\pparexpand[0]$ 
and $\nexpand[0]$ all independent of $\phi$. 

Taking account of this, the first-order equation 
of continuity is:
\eqb
\omega\frac{\partial}{\partial\phi}\left(
\nexpand[0]\Deltaexpand[1]+
\nexpand[1]\Deltaexpand[0]
\right)
+\frac{1}{{\rho}^2}\frac{\partial}{\partial {\rho}}\left({\rho}^2\nexpand[0]\pparexpand[0]\right)
+\frac{1}{\rho\sin\theta}\frac{\partial}{\partial {\theta}}\left(\sin\theta \nexpand[0]\pthetaexpand[0]\right)
+\frac{1}{\rho\sin\theta}\frac{\partial}{\partial {\varphi}}\left(\nexpand[0]\pphiexpand[0]\right)
&=&0 
\label{contone} 
\enspace,
\eqe
Faraday's and Amp\`ere's laws are:
\eqb
-\frac{\omega}{\betaw}\frac{\partial \Eexpand[1]}{\partial\phi}
-i\omega\frac{\partial \Bexpand[1]}{\partial\phi}
+
\frac{1}{{\rho}}\frac{\partial}{\partial {\rho}}\left({\rho}\Eexpand[0]\right)&=&0 
\label{appfaraday} \\
-\frac{\omega}{\betaw}\frac{\partial \Bexpand[1]}{\partial\phi}
+i\omega\frac{\partial \Eexpand[1]}{\partial\phi}
+i8\pi e\left(\nexpand[0]\pperpexpand[1]+\nexpand[1]\pperpexpand[0]\right)
&&\nonumber\\
+\frac{1}{{\rho}}\frac{\partial}{\partial {\rho}}({\rho}\Bexpand[0])
-i\frac{M}{{\rho}}8\pi e\nexpand[0]\pperpexpand[0]&=&0 
\enspace,
\label{appampere} 
\eqe
and momentum/energy equations give:
\eqb
\omega\Deltaexpand[0]\frac{\partial \pparexpand[1]}{\partial\phi}
+\frac{e}{m}\textrm{Im}\left(\pperpexpand[0]\Bexpand[1]^*
+\pperpexpand[1]\Bexpand[0]^*\right)
&&\nonumber\\
+\pparexpand[0]\frac{\partial \pparexpand[0]}{\partial {\rho}}
-\frac{e}{m}\frac{M}{{\rho}}\textrm{Im}\left(\pperpexpand[0]\Bexpand[0]^*\right)
&=&\frac{|\pperpexpand[0]|^2}{\rho} 
\label{pparone}
\\
\omega
\Deltaexpand[0]\frac{\partial \pperpexpand[1]}{\partial\phi}
+\omega\Deltaexpand[1]\frac{\partial \pperpexpand[0]}{\partial\phi}
-\frac{e}{m}\left(\gammaexpand[0] \Eexpand[1]+\gammaexpand[1]\Eexpand[0]+i\pparexpand[1]\Bexpand[0]
+i\pparexpand[0]\Bexpand[1]\right)
&&\nonumber\\
+\pparexpand[0]\frac{\partial \pperpexpand[0]}{\partial {\rho}}
+\frac{e}{m}\frac{M}{{\rho}}\left(\gammaexpand[0] \Eexpand[0]
+i\pparexpand[0]\Bexpand[0]\right)&=&-\frac{\pparexpand[0]\pperpexpand[0]}{\rho}-i\cot\theta\frac{\pperpexpand[0]\pphiexpand[0]}{\rho}
\label{pperpone}
\\
\omega
\Deltaexpand[0]\frac{\partial \gammaexpand[1]}{\partial\phi}
+\Deltaexpand[1]\frac{\partial \gammaexpand[0]}{\partial\phi}
-\frac{e}{m}\textrm{Re}\left(
\pperpexpand[0]\Eexpand[1]^*+\pperpexpand[1]\Eexpand[0]^*\right)
&&\nonumber\\
+\pparexpand[0]\frac{\partial \gammaexpand[0]}{\partial {\rho}}
+\frac{e}{m}\frac{M}{{\rho}}\textrm{Re}\left(\pperpexpand[0]\Eexpand[0]^*\right)
&=&0
\enspace.
\label{gammaone}
\eqe

The \lq\lq slow\rq\rq\ dependence of the zeroth-order quantities
on ${\rho}$ follows by eliminating secular terms in the first-order quantities,
i.e. by imposing the condition that they are periodic in $\phi$.
Equation~(\ref{contone}) can immediately
be integrated over $\phi$, yielding, when periodicity is imposed, 
\eqb
\frac{1}{{\rho}^2}\frac{\partial}{\partial {\rho}}
\left({\rho}^2\nexpand[0]\pparexpand[0]\right)&=&0 
\enspace.
\label{contfinal}
\eqe
Similarly, (\ref{appfaraday}) integrates to give
\eqb 
\frac{1}{{\rho}}\frac{\partial}{\partial {\rho}}\left({\rho}\int_0^{2\pi}
\diff\phi\,\Eexpand[0]\right)&=&0 
\enspace.
\eqe
However, this merely constrains the average components of the wave
fields, which we assume to vanish.
In order to integrate (\ref{gammaone}) and (\ref{pparone}), it is first
necessary
to use Amp\`ere's law (\ref{appampere}) to 
re-express $\pperpexpand[0]$ in the expressions
$\textrm{Im}\left(\pperpexpand[0]\Bexpand[0]^*\right)$
and $\textrm{Re}\left(\pperpexpand[0]\Eexpand[0]^*\right)$
in terms of $\Bexpand[0]$ and $\Eexpand[0]$ respectively. 
One then finds
\eqb 
\frac{\partial}{\partial {\rho}}\left[{\rho}^2\left(\pparexpand[0]\nexpand[0]
\gammaexpand[0]+\frac{\betaw\left|\Bexpand[0]\right|^2}{8\pi m}
\right)\right]&=&0
\label{energyfinal} 
\\
\frac{\partial}{\partial {\rho}}\left[{\rho}^2\left(\pparexpand[0]^2\nexpand[0]+
\left(1+\betaw^2\right)
\frac{\left|\Bexpand[0]\right|^2}{16\pi m}\right)\right]&=&\nexpand[0]|\pperpexpand[0]|^2\rho
\enspace.
\label{momentumfinal}
\eqe
Equations (\ref{contfinal}), (\ref{energyfinal}) and (\ref{momentumfinal})
suffice to determine the dependence on ${\rho}$ of the phase-averaged, 
zeroth-order variables.
Note that, to this order, $M$ does not appear; i.e., 
general relativistic effects do not enter. Furthermore, because we 
assume cold, dissipationless fluids that interact only via the 
wave fields, (\ref{energyfinal}) %and (\ref{momentumfinal}) 
simply
states the conservation of the sum of the 
zeroth-order particle and field contributions to 
the phase-averaged 
energy %and radial momentum 
flux in flat space, expressed in 
differential form. 

Integrating (\ref{contfinal}) and (\ref{energyfinal}), 
\eqb
2 m r^2 \nexpand[0]\pparexpand[0]&=&\dot{M}/\Omega_{\rm s}
\\
\mu&=&\gammaexpand[0]\left(1+\sigma\right)
\label{energyfinal2}
\eqe
and from (\ref{momentumfinal})
\eqb
\frac{d\nu}{dR}&=&\frac{R|\pperp|^2}{\mu\hat{\omega}^2}\enspace,
\eqe
where we define
\eqb
\nu&=&\pparexpand[0]\left(1+\frac{1+\betaw^2}{2\betaw^2}\sigma\right)
\enspace,
\label{momentumfinal2}
\eqe
$\dot{M}/\Omega_{\rm s}$ 
is the mass-flux per unit solid-angle, $\mu=L/\dot{M}$ is 
the mass-loading parameter, 
$\nu$ is the radial momentum flux density per unit rest
mass and the magnetisation parameter, defined as the ratio of the field
and particle terms in the energy flux density, is
\eqb
\sigma&=&\frac{\left|\Bexpand[0]/\gammaexpand[0]\right|^2}{8\pi\nexpand[0]m
}
\enspace.
\label{sigmadef}
\eqe
Using the definition of the strength parameter (\ref{azerodef}) enables
the mass-flux to be expressed in terms of $a_0$ and $\mu$, leading to 
\eqb
\pparexpand[0]&=&\frac{a_0^2}{\mu r^2\omega_{\rm p}^2}
\enspace,
\label{contfinal2}
\eqe
where $\omega_{\rm p}=\left(8\pi n e^2/m\right)^{1/2}$ 
is the \lq\lq proper\rq\rq\ plasma frequency. 

%\bibliographystyle{hapj}
%\bibliography{apj-jour,references}

\end{document}